\documentclass[runningheads]{llncs}
\usepackage[T1]{fontenc}
\usepackage{multirow}
\usepackage{booktabs}
\usepackage{amssymb}

\newcommand\ie{\textit{i.e.}}
\newcommand\eg{\textit{e.g.}}
\usepackage{algorithm}
\usepackage{algpseudocode}
\usepackage{svg}
\usepackage{xcolor}
\definecolor{red}{RGB}{230,10,30}

\usepackage{caption}
\usepackage{etoolbox}
\usepackage{amsmath}
\usepackage[capitalize]{cleveref}

\newcommand{\equalcontrib}{\textsuperscript{*}}

\usepackage{graphicx}
\usepackage[misc,geometry]{ifsym}

\begin{document}
\title{Semantic Segmentation for Preoperative Planning \\ in Transcatheter Aortic Valve Replacement}
\titlerunning{Semantic Segmentation for TAVR Preoperative Planning}
%

\author{Anonymized Authors}  
\authorrunning{Anonymized Author et al.}
\institute{Anonymized Affiliations \\
    \email{email@anonymized.com}}

\author{%
Cedric Z{\"o}llner\inst{1}\equalcontrib \and
Simon Rei{\ss}\inst{1}\equalcontrib \and
Alexander Jaus\inst{1,2}\and
Amroalalaa Sholi\inst{3}\and
Ralf~Sodian\inst{3} \and
Rainer Stiefelhagen\inst{1}}
 

\footnotetext[1]{Equal contribution}
\authorrunning{C.~Zöllner et al.} 

\institute{Karlsruhe Institute of Technology, Karlsruhe, Germany \and Helmholtz Information and Data Science School for Health, Germany \and
Department of Thoracic and Cardiovascular Surgery, Heart Centre Lahr, Germany\\
\Letter~\email{simon.reiss@kit.edu}}

\maketitle              
\begin{abstract}
    When preoperative planning for surgeries is conducted on the basis of medical images, artificial intelligence methods can support medical doctors during assessment.
    In this work, we consider medical guidelines for preoperative planning of the transcatheter aortic valve replacement (TAVR) and identify tasks, that may be supported via semantic segmentation models by making relevant anatomical structures measurable in computed tomography scans.
    We first derive fine-grained TAVR-relevant pseudo-labels from coarse-grained anatomical information, in order to train segmentation models and quantify how well they are able to find these structures in the scans.
    Furthermore, we propose an adaptation to the loss function in training these segmentation models and through this achieve a $+1.27\%$ Dice increase in performance.
    Our fine-grained TAVR-relevant pseudo-labels and the computed tomography scans we build upon are available at~\url{https://doi.org/10.5281/zenodo.16274176}.

\end{abstract}

\section{Introduction and related work}
In developed countries, aortic valve disease affects $11.7\%$ of people over the age $75$~\cite{nkomo2006burden}.
In many cases such diseases can be treated by placing an implant over the malfunctioning valve to recover its functionality.
A minimally invasive way to do this is Transcatheter Aortic Valve Replacement (TAVR).
Ahead of TAVR, preoperative planning is done, which involves measuring a patient's anatomical features based on computed tomography (CT). 
This is structured by guidelines~\cite{vahanian20222021}, which state the most important \textit{anatomy} to be assessed: The \textit{aortic valve} and the shape and size of the \textit{annulus} has to be known for the artificial valve to stay in place and seal the aorta, 
the size of the \textit{aortic root} is measured to determine the size of the stent, the vascular access to the valve is measured, which includes looking at vessels the catheter passes through, \ie, \textit{aorta}, \textit{femoral-} and \textit{iliac arteries}.
Further, surgeons focus on calcification in the above anatomy,~\ie, the aortic valve, left ventricular outflow tract and the whole vascular system.

{
\captionsetup{belowskip=-15pt}
\renewcommand{\arraystretch}{0.8}
\begin{table}[t]
\setlength{\tabcolsep}{4.2pt}
\begin{center}\begin{minipage}[b]{1.0\linewidth}
  \centering
    \begin{tabular}{lrccccccc}
    Dataset & size & \rotatebox{90}{\tiny Aorta} & \rotatebox{90}{\tiny Left Ventricle} & \rotatebox{90}{\tiny Aortic Root} & \rotatebox{90}{\tiny Valve} & \rotatebox{90}{\tiny Annulus} & \rotatebox{90}{\tiny Iliac Artery Left} & \rotatebox{90}{\tiny Iliac Artery Right }\\
    \midrule
    TotalSegmentator \cite{wasserthal_totalsegmentator_2023} & $1,228$ & \checkmark & \checkmark & -- & -- & -- & \checkmark & \checkmark  \\
    
    MM-WHS \cite{zhuang_evaluation_2019} & $60$ & \checkmark & \checkmark & -- & -- & -- & -- & -- \\
    
    SegTHOR \cite{lambert_segthor_2020} & $60$ & \checkmark & -- & -- & -- & -- & -- & -- \\
    
    DAP Atlas \cite{jaus_towards_2023} & $533$ & \checkmark & \checkmark & -- & -- & -- & \checkmark & \checkmark \\
    
    \midrule
    
    Our extension of~\cite{wasserthal_totalsegmentator_2023} & 578 & \checkmark & \checkmark & \checkmark & \checkmark & \checkmark & \checkmark & \checkmark \\
    \bottomrule
    \\
    \end{tabular}
  \centerline{(a) Table of public datasets with TAVR-related CT annotations}\medskip
\end{minipage}
\begin{minipage}[b]{1.0\linewidth}
  \centering
    \includegraphics[width=\linewidth]{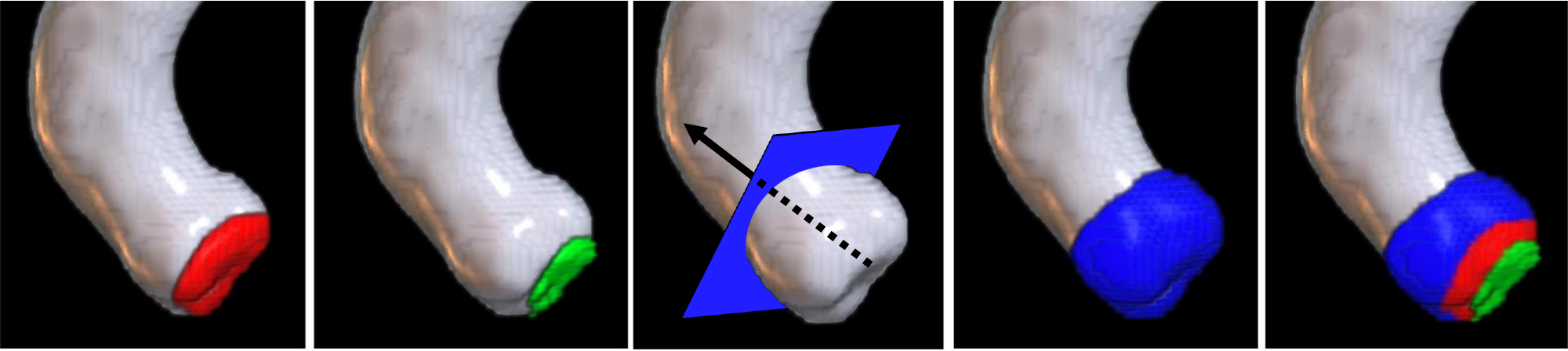}
  \centerline{(b) Additional extracted TAVR-related anatomical structures}\medskip
\end{minipage}
\end{center}

\caption{(a) Publicly available heart-related CT datasets with TAVR-relevant anatomy classes. (b) Label of \textit{aorta} (grey), of \textit{valve} (red), \textit{annulus} (green), and \textit{aortic root} (blue).}
\label{tab:table_heart_datasets}
\end{table}
}

Advancements in biomedical image segmentation~\cite{hatamizadeh_swin_2022,milletari_v-net_2016,ronneberger_u-net_2015} enable precise delineation of semantic structures in images.
While segmentation models can be utilized to measure structures in images, such as human anatomy in CTs, they require sufficiently large human-labeled training datasets to function.
In~\Cref{tab:table_heart_datasets}~(a) we show publicly available heart-related CT datasets and the present anatomy labels relevant in TAVR preoperative planning.
Some relevant anatomy is covered, some is missing and thus support via trained models in preoperative planning would be incomplete.
While prior works propose innovative approaches for TAVR surgeries~\cite{lessmann2017automatic,almar2024beyond,pak2020efficient,zheng2010automatic}, due to development on in-house data a broader investigation or comparison of such techniques is not possible.
In this work, we investigate to what extent existing public datasets can be enriched with additional TAVR relevant anatomy pseudo-labels and explore whether this anatomy can be learned by a single segmentation model to build a basis for preoperative planning assistance.
To foster exploration, we will make the enriched dataset publicly available. 
To get a fuller picture how suitable segmentation models are  for this task, we investigate the efficacy of different model architectures~\cite{hatamizadeh_swin_2022,milletari_v-net_2016,ronneberger_u-net_2015} and loss functions~\cite{lin_focal_2018,kirchhoff_skeleton_2024}.
Based on this, we propose a simple yet effective adaptation to the loss function, the \textit{focal skeleton recall loss} to dynamically emphasize hard to segment structures during training.
Our contributions summarize to:

\begin{itemize}
    \item Introduction of rule-based enrichment of volumetric annotations for TAVR-related anatomical structures.
    \item We make our enriched volumetric structures publicly available and benchmark the efficacy of segmentation models on this TAVR anatomy.
    \item We propose the \textit{focal skeleton recall loss}, 
    leading to the improved segmentation results of $83.2\%$ Dice for the TAVR anatomy.
\end{itemize}



\section{Methods}

Next, we show how to enrich the TotalSegmentator~\cite{wasserthal_totalsegmentator_2023} dataset with TAVR-related anatomy and then introduce 
our \textit{focal skeleton recall loss} for training segmentation models on this anatomy better.

\subsection{Dataset enrichment with TAVR-relevant anatomy}

In public datasets, some TAVR anatomy is not present, least incomplete are DAP~\cite{jaus_towards_2023} and TotalSegmentator~\cite{wasserthal_totalsegmentator_2023}.
Due to the larger size, we extend the latter with added TAVR anatomy.

\textbf{Aortic valve and annulus}
can be labeled by extracting all voxels within a certain distance to the aorta and the left heart ventricle.
We derive the \textit{valve} pseudo-label by taking the voxels of the aorta in the TotalSegmentator dataset that have a distance of three or less voxels to any left heart ventricle voxel. The valve is a curved structure since it is made of curved leaflets, thus we find a three voxel distance to ensure that the leaflets are fully covered (see~\Cref{tab:table_heart_datasets}~(b), red).
The \textit{annulus} is a thin structure defined as ring around the valve, where it is attached. We extract the annulus as every voxel of the left heart ventricle, which has a distance of one voxel to the aorta. The left heart ventricle has a flat end in the annotations and is therefore more suitable for extracting the annulus than the curved structure of the aorta. The result is a thin slice (see~\Cref{tab:table_heart_datasets}~(b), green), an area which the valve blocks when it is closed.
It has the size that the artificial valve has to have to replace the natural valve.

{

\begin{figure}[t]
    \centering
    \includegraphics[width=.75\linewidth]{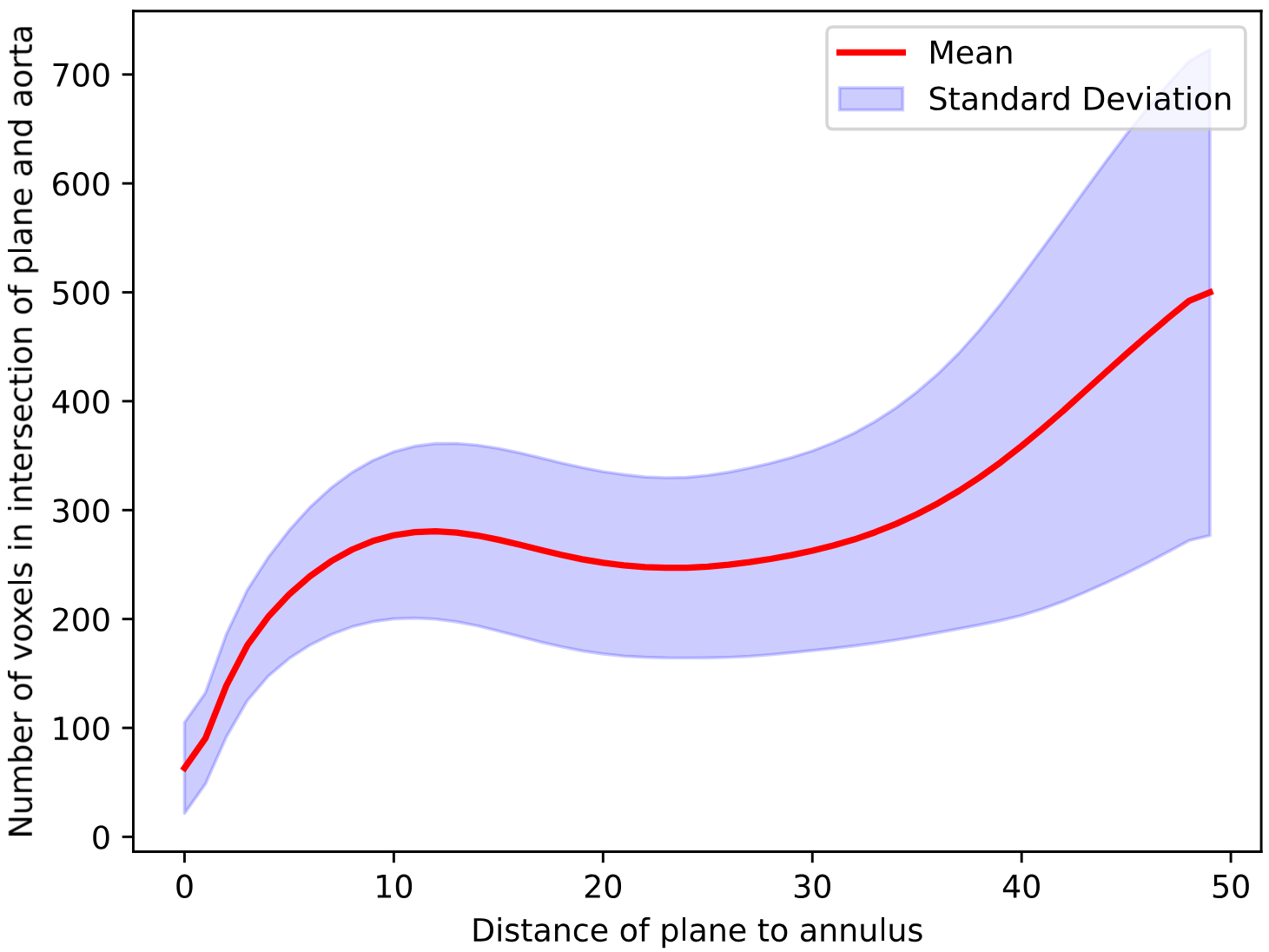}

    \caption{The number of voxels in the cross-section of the aorta over the entire dataset. The local minimum at a distance to the annulus of 25 marks the end of the aortic root.
    }
    \label{fig:RootIntersections}
\end{figure}
}

\textbf{Aortic root}
To extract information about the aortic root, a simple distance-based extraction is not possible as it is more dynamically defined as the part of the aorta
that originates from the left ventricle and has a larger cross section area than the rest of the aorta~\cite{piazza2008anatomy}.
Therefore, the end of the aortic root can be determined using the cross section area of the aorta at different distances to the ventricle, beginning with  the first cross section at the annulus.
We monitor the number of aorta voxels in the cross section starting at the annulus, where by the least squares method we fit a plane through it and successively intersect parallel planes with increasing distance with the aorta label (\Cref{tab:table_heart_datasets}~(b), third image).
These intersection areas between aorta and the plane can be plotted against the distance to the annulus in a graph.
With this, we can identify the end of the aortic root at the first local minimum after the first local maximum of the cross section area.
In~\Cref{fig:RootIntersections}, we show the aortic cross section area as compared to the annulus distance.
The local minimum after the first maximum is clearly identifiable, which is also reflected in individual CTs (increase after local minimum due to aortic arch).
For a more reliable detection of the aortic root, we use the moving average over the areas in search for the minimum.
An example for an extracted aortic root can be seen in blue in~\Cref{tab:table_heart_datasets}~(b).
    
        
        
        



The final enriched dataset now covers the TAVR anatomy \textit{aorta}, \textit{left ventricle}, \textit{aortic root}, \textit{valve}, \textit{annulus}, \textit{iliac artery left}, \textit{iliac artery right} and spans $578$ CTs. As some CTs from~\cite{wasserthal_totalsegmentator_2023} do not show TAVR-relevant anatomy we omit them.

\subsection{Focal Skeleton Recall Loss}

Our TAVR dataset contains thin, tubular structures which can be challenging to segment using neural networks.
Recently, the skeleton recall loss~\cite{kirchhoff_skeleton_2024} was introduced to address this:

\begin{equation}
    \mathcal{L}_{\text{SR}} = -\frac{1}{|C|} \sum_{c \in C} \frac{\sum_i y_{\text{skel}, i, c} \cdot p_{i, c}}{\sum_i y_{\text{skel}, i, c}}\enspace,\\[-3mm]
    \label{eq:skel}
\end{equation}
$y_{\text{skel},i,c}$ is the $i^{\text{th}}$ voxel on the thinned label, the so called skeleton of class $c$, which is computed for all classes $C$.
Further, $p_{i,c}$ denotes the sparse prediction of the segmentation model on the class-wise skeletons.
This loss was designed to improve connectivity of thin structures, by enforcing high recall on the class skeletons.
For our TAVR data we observe a severe imbalance in class distribution: most voxels are associated to the background or to the dominant aorta class, while the iliac arteries, valve or aortic root span just few voxels (\textit{cf.}~\Cref{fig:arch} first image).
Due to this, different anatomical structures are learned with varying speed as the model achieves a large loss minimization by simply recognizing \eg, the aorta. 
To heighten the influence of smaller classes dynamically, the focal loss was proposed by Lin~\textit{et al.}~\cite{lin_focal_2018}:
\begin{equation}
    \mathcal{L}_{\text{Focal}} = - (1-p_{i,c}^*)^\gamma \log(p_{i,c}^*),\enspace p_{i,c}^*= \Biggl\{
    \begin{array}{c}
    p_{i,c} \enspace \text{if} \enspace y=1 \\
    1 - p_{i,c} \enspace \text{otherwise}
  \end{array} \enspace.\\
  \label{eq:focallossformula}
\end{equation}
In this formula the class-agnostic hyper-parameter $\gamma$ is introduced to emphasize the hardness of a prediction $p_{i,c}$.
We adapt the skeleton recall loss in~\Cref{eq:skel} and take inspiration from the focal loss in~\Cref{eq:focallossformula}, yielding our \emph{focal skeleton recall loss}:

\begin{equation}
    \mathcal{L}_{\text{Focal SR}} = -\frac{1}{|C|} \sum_{c \in C} \frac{\sum_i (1 - p_{i,c})^\gamma y_{\text{skel}, i, c} \cdot p_{i, c}}{\sum_i y_{\text{skel}, i, c}}\enspace.\\
\end{equation}
Here, we apply the idea of dynamic weighting with an exponent $\gamma$ onto the recall-based loss.
Due to this, no distinction between prediction and ground-truth is made contrary to the focal formulation in~\Cref{eq:focallossformula}.
Our formulation focuses the loss more on low confidence predictions.
As the skeleton recall loss and thus also our modification, the \emph{focal skeleton recall loss} considers voxels on the class skeleton exclusively, a segmentation loss is added: $\mathcal{L_{\text{FocalSK}^\star}} = \mathcal{L}_{\text{FocalSK}} + \mathcal{L}_{\text{Focal}}$.
Here, we chose the focal loss $\mathcal{L}_{\text{Focal}}$ as auxiliary segmentation objective for dynamic weighting of the predictions in training throughout.

\section{Evaluation}

\noindent\textbf{Evaluation protocol.}
We split the enriched TAVR dataset into training / validation / test sets of size $378$/$100$/$100$.
Ablation studies are conducted on the val set and the best configurations as measured in Dice score, are evaluated on the test set once.
We segment the classes \textit{aorta}, \textit{left ventricle}, \textit{aortic root}, \textit{valve}, \textit{annulus}, \textit{iliac artery left} and \textit{iliac artery right}.

\noindent\textbf{Implementation details.}
All models were implemented in MonAI~\cite{cardoso2022monai} and trained on a NVIDIA GeForce RTX 2080 Ti with 12GB of memory.
We train each model with a varying batch size of CT scans, where we extract multiple subvolumes in each training iteration depending on the memory consumption of the trained models: six  $96 \times 96 \times 96$ subvolumes from three CTs for Unet~\cite{ronneberger_u-net_2015}, 13  $64 \times 64 \times 64$ subvolumes from one CT for Vnet~\cite{milletari_v-net_2016} and four subvolumes $64 \times 64 \times 64$ from one CT for Swin UNETR~\cite{hatamizadeh_swin_2022}.
Our baseline loss function is a combination of the Dice loss and cross-entropy loss (DiceCE), which is weighted with factors of $0.25$ and $0.75$, respectively.
When focal loss or our \emph{focal skeleton recall loss} is in use, we set $\gamma = 2$ as in~\cite{lin_focal_2018}.
We train for $350$ epochs.

{
\setlength{\tabcolsep}{5pt}
\begin{table}[t]
\centering
\scriptsize
\begin{tabular}{cccc|ccccccc|c}
\toprule
\rotatebox{90}{DiceCE} & \rotatebox{90}{$\mathcal{L}_{\text{Focal}}$} & \rotatebox{90}{$\mathcal{L}_{\text{SR}}$} & \rotatebox{90}{$\mathcal{L_{\text{Focal SR}}}$} & \rotatebox{90}{Aorta} & \rotatebox{90}{left Ventr.} & \rotatebox{90}{Aortic Root} & \rotatebox{90}{Valve} & \rotatebox{90}{Annulus} & \rotatebox{90}{I. A. left} & \rotatebox{90}{I. A. right} & \rotatebox{90}{Mean} \\

\midrule

\checkmark & -- & -- & -- & 95.09 & 94.23 & 84.55 & 81.27 & 50.50 & 83.82 & 84.95 & 81.78 \\

-- & \checkmark & -- & -- & \textbf{95.99} & \textbf{95.33} & 85.56 & \textbf{83.09} & 53.70 & \textbf{85.91} & 83.59 & 83.08\\
\checkmark & -- & \checkmark & -- & 94.83 & 94.31 & 84.11 & 80.84 & 53.32 & 83.00 & 84.05 & 81.89 \\
-- & \checkmark & \checkmark & -- & 95.58 & 95.28 & 84.64 & 82.56 & \textbf{55.06} & 82.69 & 85.63 & 82.92 \\
-- & \checkmark & -- & \checkmark & 95.95 & 95.13 & \textbf{85.73} & 82.93 & 54.39 & 85.54 & \textbf{87.30} & \textbf{83.50}\\

\bottomrule

\end{tabular}

\caption{Performance of Swin UNETR with different loss functions measured in Dice score on the validation set.}
\label{tab:swin_losses_val}
\end{table}
}

\noindent\textbf{Loss ablation study.}
In~\Cref{tab:swin_losses_val}, we methodically analyze the effect of different loss combinations on the validation score when using the Swin UNETR architecture.
First, when using the standard DiceCE loss setup, we observe a mean Dice score of $81.78\%$ across all anatomical structures.
Next, we utilize the focal loss by Lin \textit{et al.}~\cite{lin_focal_2018} in training, where we can observe an increase in Dice score of $+1.3\%$, indicating, that a dynamic weighting of the loss to focus on incorrectly predicted regions helps.
The skeleton recall loss~\cite{kirchhoff_skeleton_2024} is originally used in conjunction with a standard segmentation loss, so we combine it with the DiceCE loss in row three of~\Cref{tab:swin_losses_val}.
While this improves the mean Dice score by $+0.2\%$, out-of-the-box, the skeleton recall loss has only a slight effect.
In line two, we found focal loss to be a better choice as compared to DiceCE, so next, we combine the skeleton recall loss with the focal loss in row four.
While this combination does perform better than skeleton recall with DiceCE, compared to Swin UNETR trained with focal loss exclusively (line two), the skeleton recall loss has a harmful effect of $-0.16\%$.
With this insight, we move to our adaptation to the skeleton loss, the \emph{focal skeleton recall loss}, which we combine with the focal loss as in $\mathcal{L_{\text{FocalSK}^\star}}$.
Our loss function produces the best results at $83.5\%$ Dice, with an improvement of $+1.72\%$ over using DiceCE and $+0.42\%$ over the competitive focal loss. 

{
\setlength{\tabcolsep}{5pt}
\renewcommand{\arraystretch}{1.4}
\begin{table*}[t]
\begin{center}

\begin{tabular}{l|cc|cc|cc}
\toprule
 & \multicolumn{2}{c|}{Unet} & \multicolumn{2}{c|}{Vnet} & \multicolumn{2}{c}{Swin UNETR} \\ 
Target  & DiceCE & $\mathcal{L_{\text{FocalSK}^{\star}}}$ & DiceCE & $\mathcal{L_{\text{FocalSK}^{\star}}}$ & DiceCE & $\mathcal{L_{\text{FocalSK}^{\star}}}$ \\
\midrule
Aorta Dice           & 96.22 & 96.94 & 95.40 & 95.72 & 96.21 & \textbf{97.00} \\
Aorta IoU            & 92.81 & 94.12 & 91.84 & 92.24 & 93.12 & \textbf{94.41} \\
Left Ventr. Dice     & 94.31 & \textbf{95.35} & 94.22 & 94.59 & 94.30 & 95.17 \\
Left Ventr. IoU      & 89.34 & \textbf{91.19} & 89.25 & 90.03 & 89.34 & 90.89 \\
Aortic Root Dice     & 83.71 & 85.24 & 85.26 & 84.93 & 84.63 & \textbf{86.44} \\
Aortic Root IoU      & 73.89 & 75.72 & 76.01 & 75.36 & 74.89 & \textbf{77.34} \\
Valve Dice           & 81.97 & 82.73 & 81.75 & 80.97 & 81.78 & \textbf{83.28} \\
Valve IoU            & 70.97 & 71.56 & 70.60 & 69.58 & 70.29 & \textbf{72.23} \\
Annulus Dice         & \textbf{56.13} & 52.97 & 54.16 & 48.27 & 53.80 & 55.23 \\
Annulus IoU          & \textbf{41.50} & 38.29 & 39.78 & 34.45 & 39.12 & 40.79 \\
I.\,A.\,left Dice    & 79.79 & 79.39 & 81.06 & \textbf{81.38} & 79.81 & 81.14 \\
I.\,A.\,left IoU     & 69.78 & 69.18 & \textbf{72.38} & 72.28 & 69.87 & 72.17 \\
I.\,A.\,right Dice   & 78.09 & 80.02 & 82.48 & 78.93 & 81.71 & \textbf{83.43} \\
I.\,A.\,right IoU    & 67.09 & 69.02 & 72.47 & 68.92 & 71.08 & \textbf{73.93} \\
\midrule
Mean Dice            & 81.89 & 82.09 & 81.93 & 80.63 & 81.86 & \textbf{83.20} \\
Mean IoU             & 72.91 & 73.32 & 73.19 & 71.94 & 72.86 & \textbf{74.79} \\
\bottomrule
\end{tabular}
\end{center}

\caption{Performance of different segmentation architectures trained with Dice and cross-entropy loss as well as our \textit{focal skeleton recall loss} $\mathcal{L_{\text{FocalSK}^\star}}$. Dice score and IoU metrics, measured on the test set.}
\label{tab:architectures_losses_test}
\end{table*}
}

{
\begin{figure}[ht!]
    \begin{tabular}{cccc}
        Target & Unet & Vnet & Swin UNETR \\
          \includegraphics[width=0.16\columnwidth]{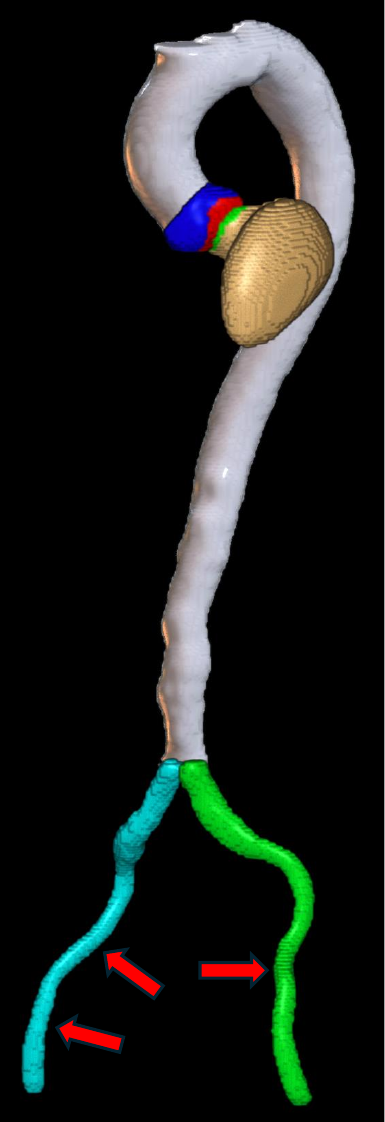} & \includegraphics[width=0.16\columnwidth]{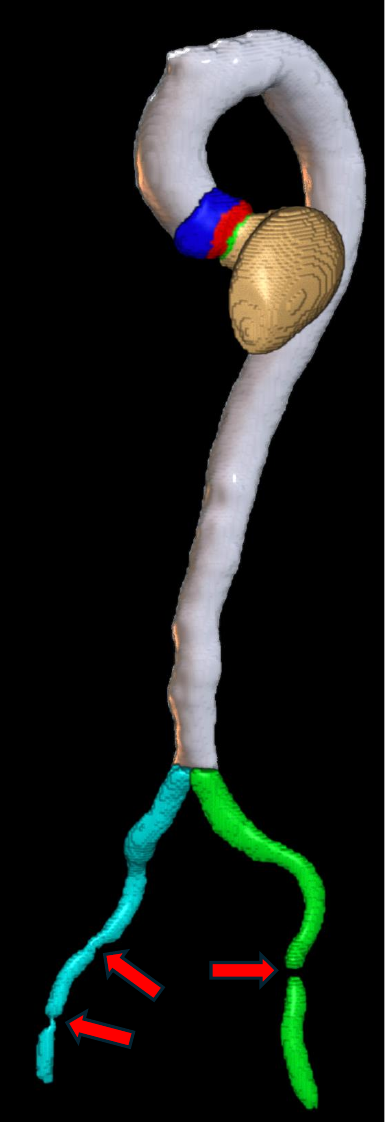} & \includegraphics[width=0.16\columnwidth]{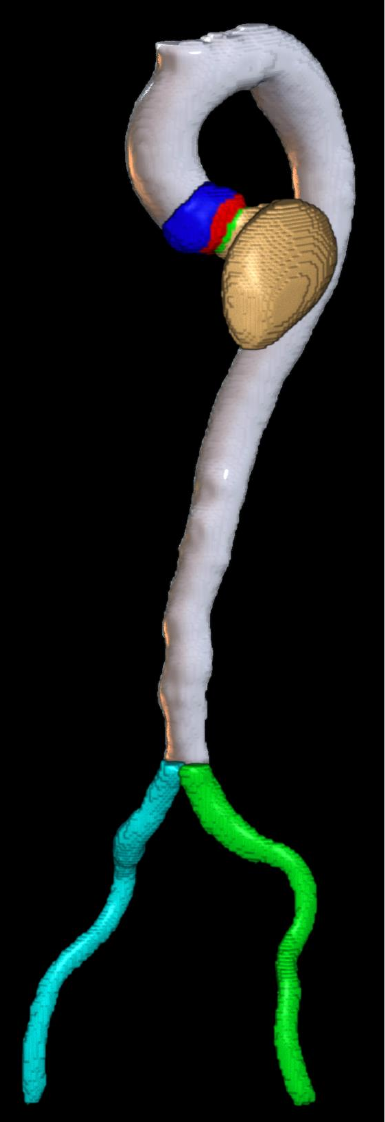} & \includegraphics[width=0.16\columnwidth]{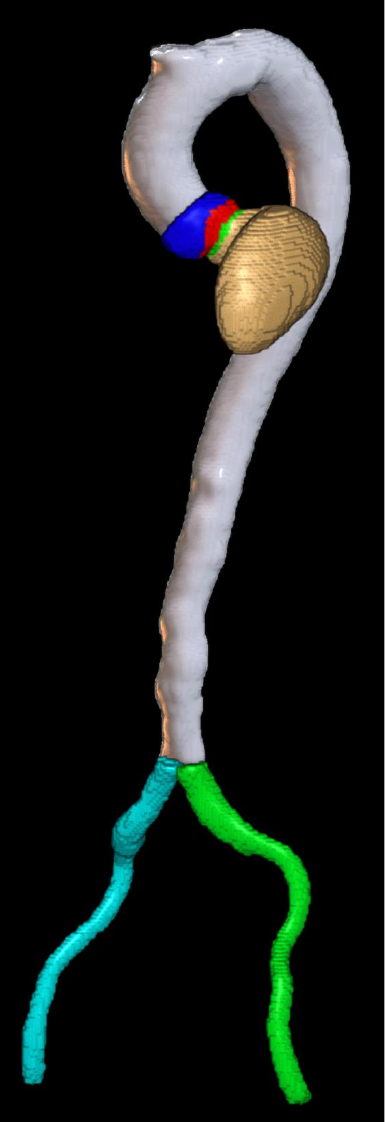} \\
         \includegraphics[width=0.16\columnwidth]{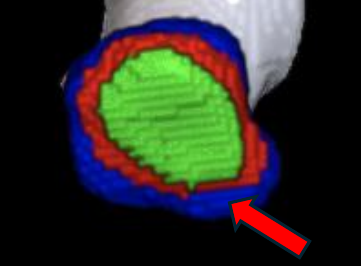} & \includegraphics[width=0.16\columnwidth]{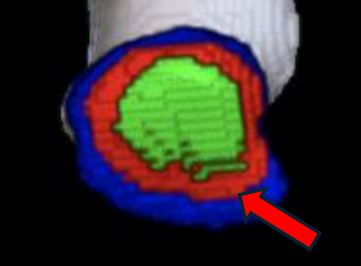} & \includegraphics[width=0.16\columnwidth]{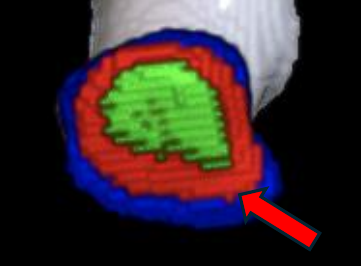} & \includegraphics[width=0.16\columnwidth]{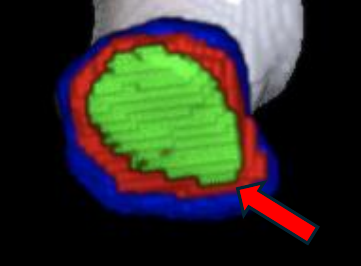}
    \end{tabular}
    \centering
    \begin{tabular}{ccc}
         Target & DiceCE & $\mathcal{L_{\text{FocalSK}^\star}}$ \\
         \includegraphics[width=0.24\columnwidth]{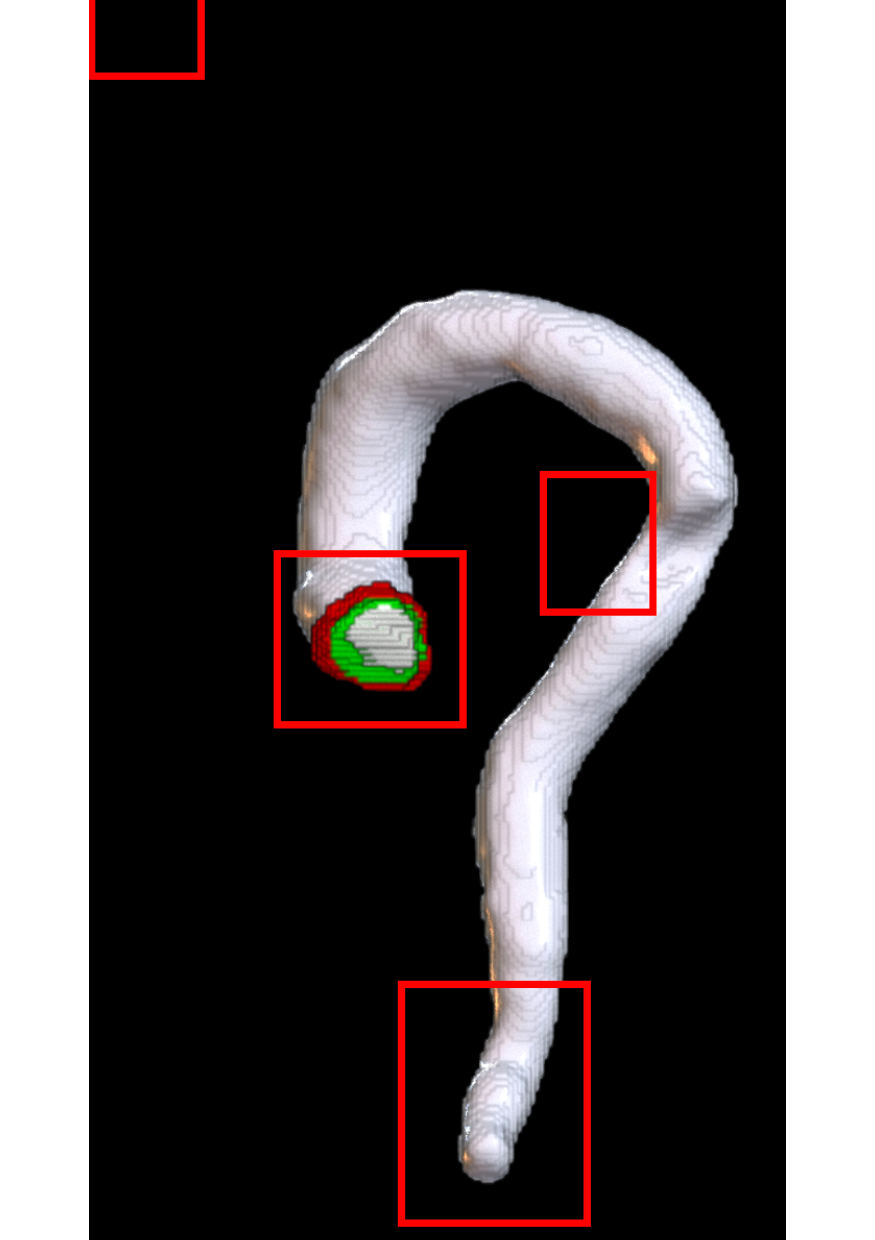} & \includegraphics[width=0.24\columnwidth]{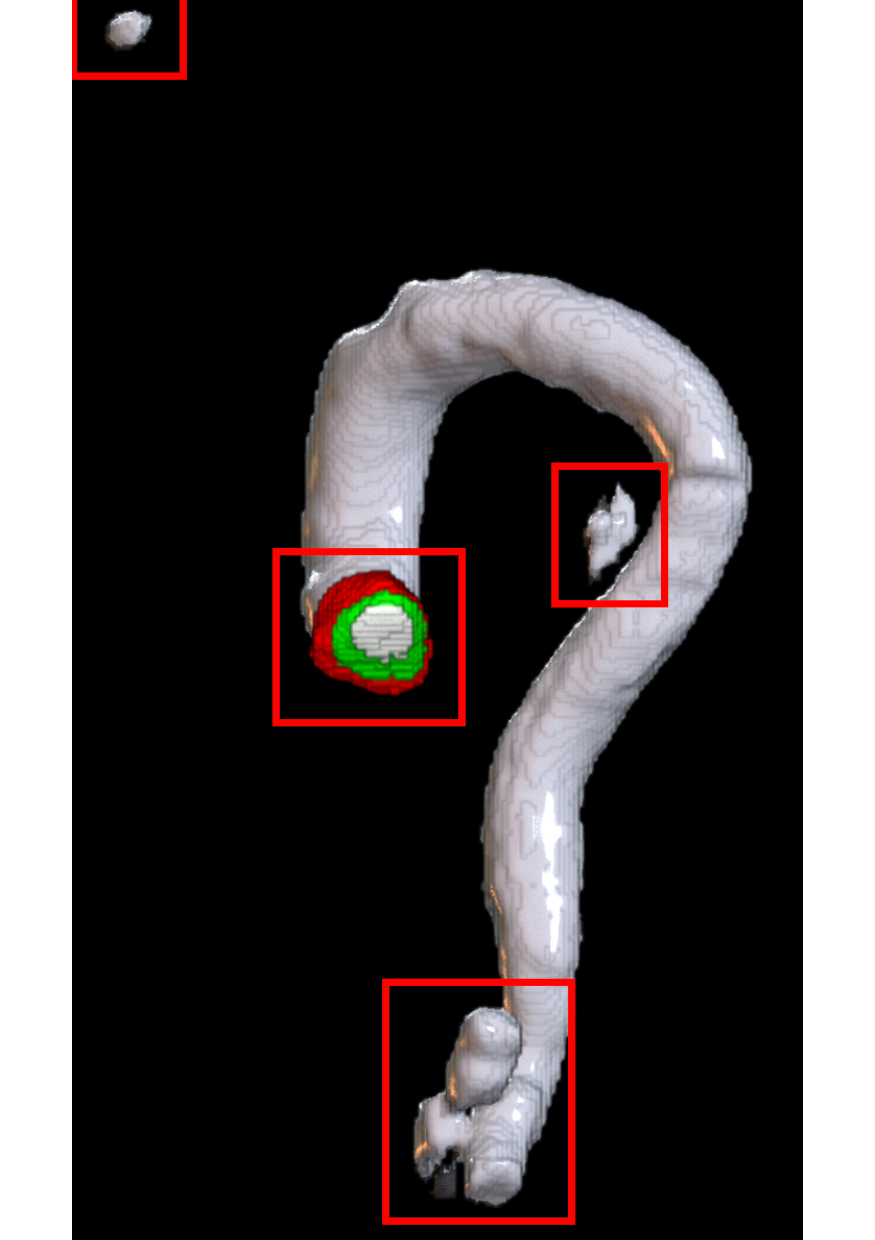} & \includegraphics[width=0.24\columnwidth]{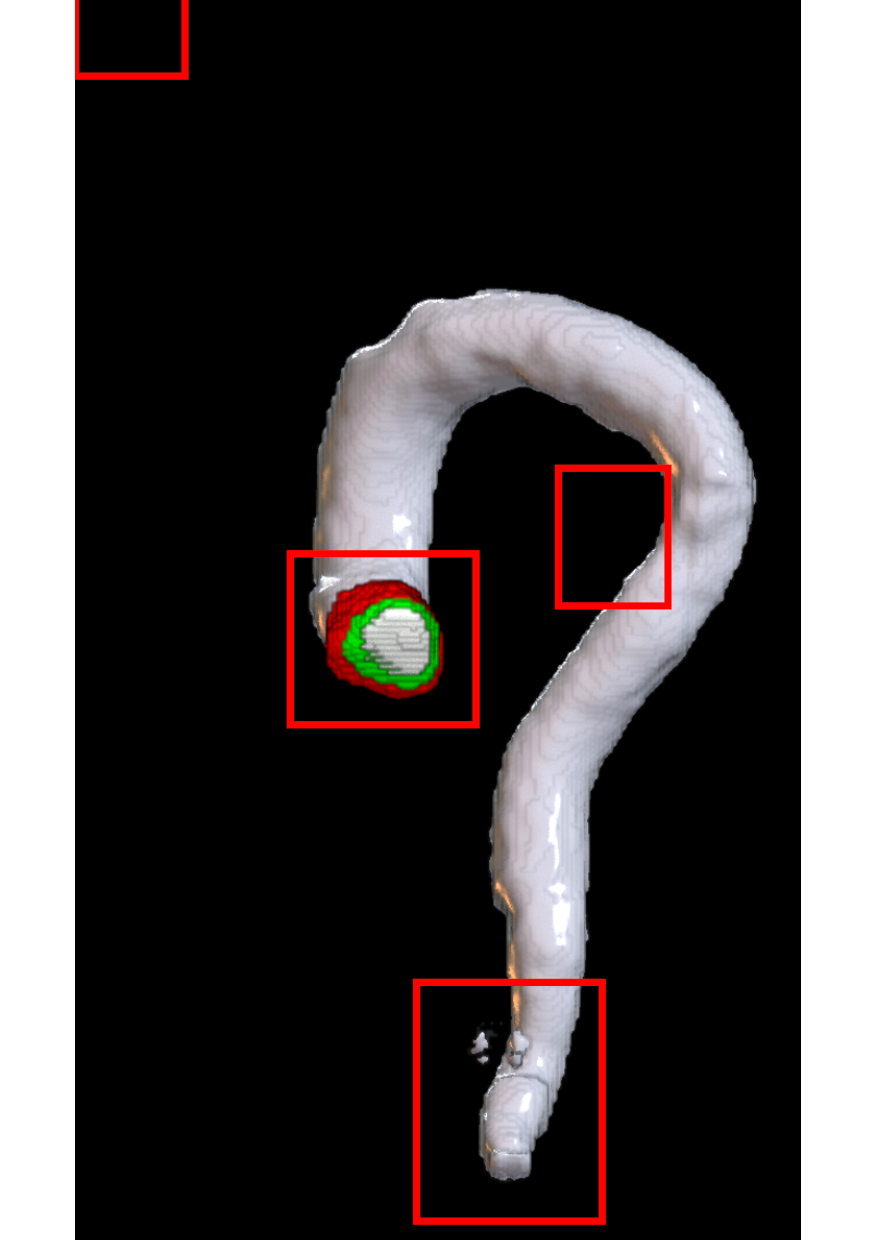}
    \end{tabular}
    \caption{Comparison between target and predictions of Unet, Vnet  and Swin UNETR, using our $\mathcal{L}_{\text{FocalSK}}$ in the first two rows, bottom row shows segmentation difference of Swin UNETR trained with DiceCE and $\mathcal{L}_{\text{FocalSK}}$ loss functions. Iliac arteries (turquoise, green), aorta (grey), aortic root (blue), valve (red), annulus (light green).}
    \label{fig:arch}
\end{figure}
}

\noindent\textbf{Quantitative results.}
Next, we focus our attention on the evaluation of different model architectures trained with DiceCE loss and our loss variation.
Here, we test all three architectures, the Unet, Vnet and Swin UNETR.
We report the Dice scores and mean Intersection over Union (IoU) metrics on the test set in~\Cref{tab:architectures_losses_test}.
First, when training the three architectures with the standard DiceCE loss, the performance difference between them is negligible, with the maximal difference in Dice spanning merely $0.07\%$, \ie, Unet with $81.89\%$,  Vnet with $81.93\%$ and Swin UNETR with $81.86\%$ Dice, respectively.
This seems to imply, that in this comparison, the specific architecture used on the dataset, is not an important factor.
Where the architecture starts to lead to differing results is when we train with our modified \emph{focal skeleton recall loss}.
While Vnet does not profit from the integration of the loss, Unet and especially Swin UNETR behave quite differently.
Unet's performance improves by a moderate $+0.2\%$, while Swin UNETR improves by a considerable $+1.34\%$ mean Dice, which is the best score achieved among all our experiments.
Especially, the left- and right iliac artery with their tubular structure seem to profit from the \emph{focal skeleton recall loss}, where their Dice scores improve by $+1.33\%$, $+1.72\%$ and their IoU improve by $+2.3\%$, $+2.85\%$, respectively when changing DiceCE to \emph{focal skeleton recall loss} using Swin UNETR.
Further, we see the Dice and IoU scores for the aorta ($+0.79\%$/$+1.29\%$), left ventricle ($+0.87\%$/$+1.55\%$), aortic root ($+1.81\%$/$+2.45\%$), valve ($+1.5\%$/$+1.94\%$) and annulus ($+1.43\%$/$+1.67\%$) are all affected favourably when shifting from DiceCE to our loss function.

\noindent\textbf{Qualitative results.}
In order to assess the difference in volumetric segmentation efficacy of the presented segmentation models, we present some segmentation predictions in~\Cref{fig:arch}.
There, next to the target segmentation, in the upper row, the predictions of all TAVI-relevant anatomy via Unet, Vnet and Swin UNETR, trained with our \emph{focal skeleton recall loss} are shown.
The overall segmentation of the aorta (grey) is captured well by all models, although, when looking into the segmentation of the right and left iliac arteries (turquoise and green, indicated with red arrows), differences become more evident.
While Unet has visible missed regions in the iliac arteries, both Vnet and Swin UNETR produce smooth, continuously connected tubes, coherent to the target segmentation.
In the second row of images, errors in the segmentation of the valve (red) and annulus (light green) structures become evident.
Specifically, Unet and Vnet produce speckled predictions while Swin UNETR leads to the most coherent segmentation. 
In the last row, we further take a look into segmentations that are produced by a Swin UNETR with either the DiceCE loss and our \emph{skeleton focal loss}.
The Swin UNETR trained with DiceCE loss produces unconnected aorta segments throughout multiple locations in the CTs.
As seen by the red boxes, training with our $\mathcal{L_{\text{FocalSK}^\star}}$ nicely eliminates these unconnected, false segments.

\section{Discussion and Conclusion}

We proposed a pathway to get hold of more pseudo-labels of TAVR-relevant anatomy in order to train neural segmentation models.
Via rule-based processing, we derived additional information on the \textit{aortic root}, \textit{valve} and \textit{annulus} and proceeded to train segmentation models on TAVR-relevant structures leading to a quite high score of $83.5\%$ in Dice, which enables measurements based on segmentation in an automatic fashion.
For some structures it is not yet possible to coherently segment them, such as the \textit{annulus}, and for some structures higher segmentation accuracy might be needed due to their criticality in TAVR surgery such as the \textit{valve} and \textit{iliac arteries}.
One step into this direction was the proposition of our \emph{focal skeleton recall loss}, which noticeably improved the performance in both Dice and IoU by $+1.34\%$ and $+1.93\%$, respectively.
Finally, to provide a holistic assistance to surgeons in preoperative planning for TAVRs further structures stated in medical guidelines need to be quantified, such as complex calcification patterns, femoral arteries and coronary ostia locations which are crucial to mitigate risks for complications during surgery and a hightened focus on different valve types (The Sievers Classification of bicuspid aortic valves). 
Calcifications were investigated previously~\cite{lessmann2017automatic}, yet, currently no labels are publicly accessible. Thus, some structures require data annotation by medical professionals in the future for deep learning-assisted TAVR planning. \\


\noindent\textbf{Acknowledgements}:
This work was supported by funding from the pilot program Core-Informatics of the Helmholtz Association (HGF) and the research school “HIDSS4Health – Helmholtz Information and Data Science School for Health”.

\bibliographystyle{splncs04}
\bibliography{refs}

\end{document}